\documentclass{iopart}
\usepackage{cite}
\usepackage{iopams}
\usepackage{graphicx}
\usepackage{calc}
\usepackage{amssymb}
\usepackage{color}
\usepackage[pdftex,dvipsnames,usenames]{xcolor}
\usepackage[colorlinks=true,urlcolor=blue,citecolor=blue,linkcolor=blue]{hyperref}

\newcommand{\ket}[1]{\left|#1\right\rangle}
\newcommand{\bra}[1]{\left\langle#1\right|}
\newcommand{\D}{\mbox{\rm d}}
\newcommand{\Hermite}{\mbox{\rm H}}
\renewcommand{\Re}{\mbox{\rm Re}}

\begin{document}
	\title{Dual form of the phase-space classical simulation problem in quantum optics}
	
	\author{A. A. Semenov$^{1,2,3}$ and A. B. Klimov$^4$}
	\address{$^1$Bogolyubov Institute for Theoretical Physics, NAS of Ukraine, Vul.~Metrologichna 14b, 03143 Kyiv, Ukraine}
	\address{$^2$Kyiv Academic University, Blvd. Vernadskogo  36, 03142  Kyiv, Ukraine}
	\address{$^3$Institute of Physics, NAS of Ukraine, Prospect Nauky 46, 03028 Kyiv, Ukraine}
	\address{$^4$Departamento de F\'isica, Universidad de Guadalajara, 44420 Guadalajara, Jalisco, Mexico}

\begin{abstract}
	In quantum optics, nonclassicality of quantum states is commonly associated with negativities of phase-space quasiprobability distributions.
	We argue that the impossibility of any classical simulations with phase-space functions is a necessary and sufficient condition of nonclassicality.
	The problem of such phase-space classical simulations for particular measurement schemes is analysed in the framework of Einstein-Podolsky-Rosen-Bell's principles of physical reality.
	The dual form of this problem results in an analogue of Bell inequalities.
	Their violations imply the impossibility of phase-space classical simulations and, as a consequence, nonclassicality of quantum states. 
	We apply this technique to emblematic optical measurements such as photocounting, including the cases of realistic photon-number resolution and homodyne detection in unbalanced, balanced, and eight-port configurations.
\end{abstract}

	\maketitle


\section{Introduction}

	Traditionally, classical and quantum physics are described by using distinct mathematical tools. Formulating these theories on the same language may help to understand both their differences and similarities.
	This idea lies in the foundation of the phase-space representation of quantum physics initialized in the pioneering works of Weyl \cite{weyl27}, Wigner \cite{wigner32}, Groenewold \cite{Groenewold1946}, and Moyal \cite{moyal1949}.  
	In this approach one of the key differences between quantum and classical physics consists in the appearance of negative values of phase-space probability distributions that represent states of the system.
	The impossibility to interpret negative probabilities in the framework of classical theories is often referred to as nonclassicality.
	
	The phase-space treatment of nonclassicality is commonly restricted to bosonic systems and is widely used in quantum optics. 
	The Weyl-Wigner-Moyal-Groenewold map is not a unique phase-space representation for bosonic quantum systems.
	In particular, there exists a family of $s$-parametrized Cahill-Glauber quasiprobability distributions $P(\alpha;s)$ \cite{cahill69,cahill69a} with $s\in[-1,1]$, where the
	Wigner function corresponds to the special choice $s=0$.
	The Husimi-Kano Q function \cite{husimi40,kano65} and the Glauber-Sudarshan P function \cite{glauber63c,sudarshan63} belong to the s-parametrized family with $s=-1$ and $s=1$, respectively.
		
	In quantum optics, the standard analysis of a quantum state is based on the decomposition of the density operator $\hat{\rho}$ over the coherent state projectors $\ket{\alpha}\bra{\alpha}$,
	\begin{equation}
	\hat{\rho}=\int_{\mathbb{C}}\D^{2}\alpha \,P(\alpha ) \ket{\alpha}\bra{\alpha},\label{Eq:PFunc}
	\end{equation}   
	where  $P(\alpha)$ is the Glauber-Sudarshan P function.
	If $P(\alpha)\geq0$, then the quantum state is a classical statistical mixture of coherent states; otherwise, the state is considered nonclassical \cite{titulaer65, mandel86,mandel_book,vogel_book,agarwal_book,sperling2018a,sperling2018b,sperling2020}.
	From this point of view, phenomena such as quadrature \cite{Wu1986,Wu1987, Vahlbruch, Stoler1970, Stoler1971} and photon-number \cite{mandel79} squeezing are nonclassical.
	It is worth noting that the P function is not positive semidefinite if $P(\alpha;s)\ngeq0$ for any $s<1$.
	Tests and quantifications of this type of nonclassicality require involved methods (see, e.g., references \cite{Hillery1987,Lee1991,vogel00,richter02,Asboth2005,kiesel10, rivas2009,Miranowicz2015a,Miranowicz2015b,bohmann2020,bohmann2020b}) since P functions are typically highly singular. 
	
\begin{figure}[th]
		\begin{center}
			\includegraphics[width=0.6\linewidth]{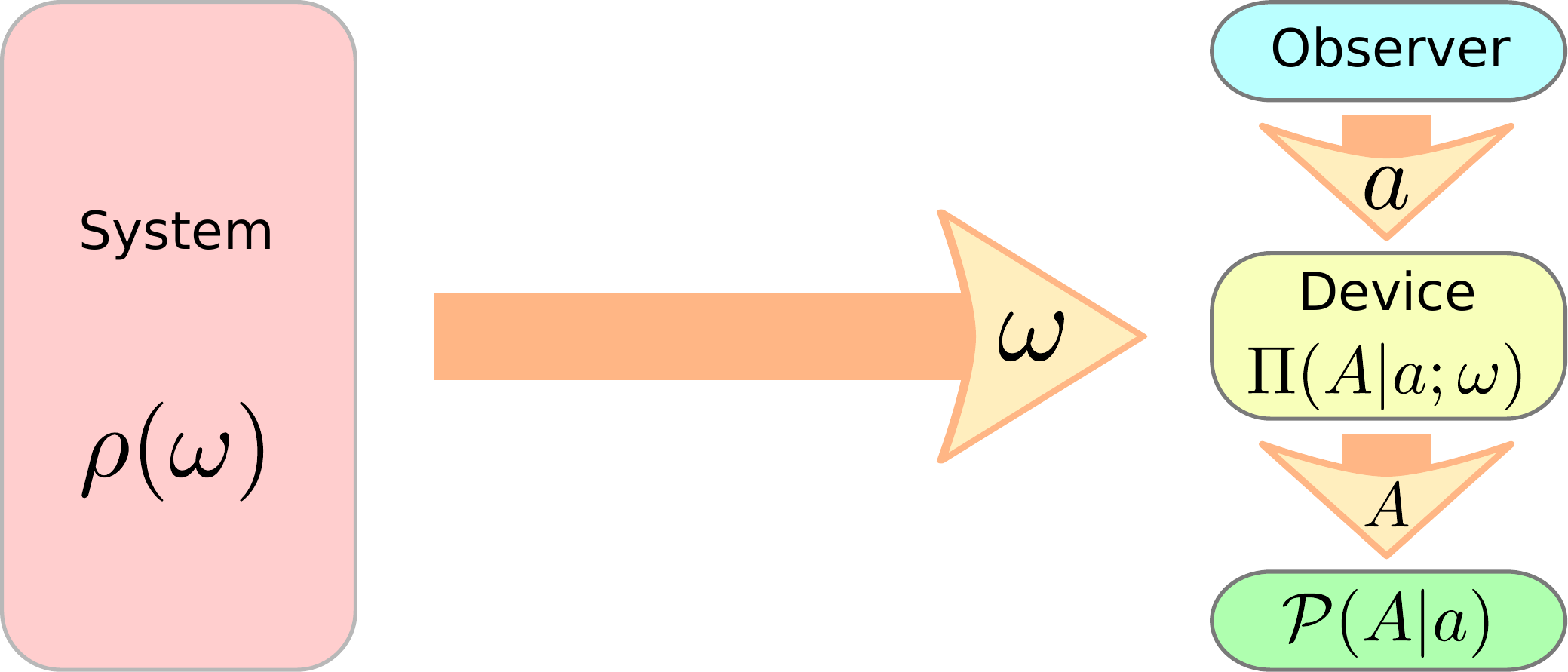}
		\end{center}
		\caption{The idea of classical simulation of quantum measurements is depicted. 
			See the main text for details.}
		\label{Fig:ClassState}
	\end{figure}

    The possibility of classical simulations of quantum measurements is another fundamental issue for which non-negative probabilities play a crucial role.
	Such simulations are convenient to consider in the framework of Einstein-Podlsky-Rosen-Bell's principles of physical reality \cite{einstein35, bell64}.
	Specifically, such classical simulations, see figure~\ref{Fig:ClassState}, assume that any physical state can be characterized by a probability distribution $\rho (\omega )\geq 0$ of some parameters $\omega$. 
	Independently on the values of $\omega$, an observer chooses a measurement-device setting $a$. 
	Measurement outcomes $A$ are described by a response function $\Pi (A|a;\omega)\geq 0$ interpreted as the probability distribution to get  the value $A$ given $a$ and $\omega$.
	Thus, the probability distribution $\mathcal{P}\left( A|a\right) $ to get an outcome $A$ given the experimental setting $a$ is computed by averaging $\Pi (A|a;\omega )$ over the parameter space $\Omega$:
	\begin{equation}\label{Eq:EPRB}
	\mathcal{P}\left(A|a\right)=\int_\Omega \D\omega\,\Pi\left(A|a;\omega\right)\rho\left(\omega\right).
	\end{equation}
	The device settings $a$ and the measurement outcomes $A$ are elements of some sets $\mathcal{O}$ and $\mathcal{I}$, respectively.
	
	In the scenario under consideration, there always exist non-negative $\rho\left( \omega \right)$ and $\Pi (A|a;\omega )$ for any non-negative $\mathcal{P}(A|a)$. 
	This implies a possibility of classical simulations for any set of quantum measurements by which we mean that $\mathcal{P}\left(A|a\right)$ can be reconstructed by sampling over  the parameters $\omega$ and measurement outcomes $A$.
	In general, such simulations cannot be obtained by injective affine maps from the space of operators into the space of functions on $\Omega$ \cite{ferrie2010}.
	At the same time, the lack of affine map does not imply the possibility of classical simulations for a given quantum state.
	
	In general, classical simulations of quantum measurements may  not be possible for a restricted set of response functions.
	For instance, in the framework of local realism \cite{bell64,Brunner2014} one deals with two (or more) spatially-separated parties with measurement outcomes $A$ and $B$ given the settings $a$ and $b$, respectively. 
    Locality of measurements means that we consider only a factorized form of response functions.
	The search for non-negative functions in the right-hand side of the equation
	    \begin{equation}
	        \mathcal{P}\left( A,B|a,b\right) =\int_{\Omega }\D\omega \,\Pi _{\mathrm{A}}\left(A|a;\omega \right) \Pi _{\mathrm{B}}\left( B|b;\omega \right) \rho \left(\omega \right)
	        \label{Eq:EPRB-2}
	    \end{equation}
	can be recast \cite{fine82,kaszlikowski2000,abramsky2011} as the primal form of a linear-programming problem \cite{Brunner2014,boyd_book}.
	The dual form of this problem, expressed as a set of inequalities for $\mathcal{P}\left( A,B|a,b\right)$, is known as Bell inequalities. 
	Violation of any one of inequality implies the inexistence of non-negative solutions of the primal form of this problem. 
	This means that the probability distributions $\mathcal{P}\left( A,B|a,b\right)$ violate local realism.
	
	Another type of restrictions for the response functions is closely related to the notion of nonclassicality in quantum optics.
	In this context, non-negative phase-space quasi-probability distributions and phase-space symbols of the positive operator-valued measures (POVM) can be used for effective classical simulations of quantum measurements  \cite{rahimi-keshari16}.
	In order to study such an important type of simulations, we restrict the form of response functions by phase-space symbols of the POVM, when phase space is chosen as the parameter space   $\Omega$.
	The ordering parameter $s$ that defines the phase-space representation should be chosen such that these symbols are non-negative. 
	Such a particular type of simulations with non-negative phase-space functions will be referred to as phase-space classical simulations\footnote{Another type of phase-space classical simulations were discussed in references \cite{opanchuk2014,reid2014,drummond2021,rosales2014,drummond2014}.}.
	
	In this paper we argue that	the impossibility of phase-space classical simulations is a sufficient condition for nonclassicality in quantum optics.     
	We set up the problem of phase-space classical simulations and formulate its dual form using the same methodology as used for testing local realism.
	This dual form results in an analogue of Bell inequalities.   
	Their violations imply the impossibility of phase-space classical simulations and, as a consequence, nonclassicality of the considered state. 
	
	Local realism is a device-independent concept since its violation can be tested by solely scrutinizing the measurement data without specifying the measurement device \cite{Brunner2014,Acin2007}.
	In contrast, phase-space classical simulations do require such a device specification which is encoded in the symbol of the corresponding  POVM.
	In addition, phase-space classical simulations explicitly depend on the ordering parameter $s$.
	We demonstrate that two values of this parameter play crucial roles: $s=1$ related to the Glauber-Sudarshan representation and a particular value $s=s_\mathrm{th}$ specific to each measurement procedure.
	
	The rest of the paper is organized as follows.
	In section \ref{Sec:BellIneq} we establish the methodology of our paper by reminding the reader of standard Bell inequalities as an example of the dual form of a linear-programming problem.
	In section~\ref{Sec:NonclProbl} we formulate the problem of  phase-space classical simulations and link it to nonclassicality of quantum states.
	An analogue of Bell inequalities for testing impossibility of  phase-space classical simulations and, as a consequence, nonclassicality of quantum states are presented in section \ref{Sec:DD-BI}.
	Examples of photocounting measurements, unbalanced, balanced, and eight-port homodyne detection (heterodyne detection) are presented in section~\ref{Sec:Appl}.
	Summary and concluding remarks are given in section \ref{Sec:Conclusions}.


\section{Preliminaries: Bell inequalities}
\label{Sec:BellIneq}

   In this paper we consider phase-space classical simulations and nonclassicality in quantum optics in the framework of the dual form of a linear programming problem.
   Originally, such a methodology was applied to a different fundamental issue---the problem of local realism---and led to the standard Bell inequalities \cite{Brunner2014}.
   In this section, we remind the relation between Bell inequalities and the dual form of a linear-programming problem.
   In the following sections, we will generalize this technique by adapting it to phase-space classical simulations.
   
   In the framework of local realism, the probability distributions $\mathcal{P}(A,B|a,b)$, of equation (\ref{Eq:EPRB-2}), are marginals of a  non-negative probability distribution $\mathcal{W}\big(\{A\},\{B\}|\{a\},\{b\}\big)$ of the sets of values $A$ and $B$ given all possible respective settings $\{a\}$ and $\{b\}$, \cite{fine82,kaszlikowski2000,abramsky2011}:
        \begin{equation}
			\mathcal{P}\left( A,B|a^{\prime },b^{\prime }\right) =
			\sum\limits_{\begin{array}{c}{\scriptstyle\{A|a\neq a^\prime\}}\\[-0.5ex]{\scriptstyle\{B|b\neq b^\prime\}}\end{array}}
			\mathcal{W}\big(\{A\},\{B\}|\{a\},\{b\}\big). \label{Eq:Marginal_Comp}
		\end{equation}    
   	This probability distribution is normalized as
         \begin{equation}   
            \sum\limits_{\{A\},\{B\}}\mathcal{W}\big(\{A\},\{B\}|\{a\},\{b\}\big) =1.
        \end{equation}
   	The vector form of equation (\ref{Eq:Marginal_Comp}) is 
        \begin{equation}
            \mathbf{P}=\mathcal{M}\mathbf{W},  \label{Eq:Marginal}
        \end{equation}
    where $\mathbf{P}$ and $\mathbf{W}$ are formed from all values of $\mathcal{P}\left(A,B|a,b\right) $ and $\mathcal{W}\big(\{A\},\{B\}|\{a\},\{b\}\big)$, respectively; the entries of the incidence matrix $\mathcal{M}$ are 0 and 1.
    Search for non-negative solutions $\mathbf{W}$ of equation (\ref{Eq:Marginal}) is the primal form of a typical problem of linear programming \cite{boyd_book}.

    Let us formulate the corresponding dual form of this problem \cite{Brunner2014,PCRyl2018,PCRyl2020}.
    Consider a vector $\mathbf{f}$ of the same dimension as $\mathbf{W}$.
    Since the components of $\mathbf{W}$ are non-negative and sum to unity, we have $\mathbf{f}^{\mathrm{T}}\mathbf{W}\leq \sup_{i}f_{i}\equiv\sup \mathbf{f}$. 
    Taking a scalar product of both sides of equation~(\ref{Eq:Marginal}) with an arbitrary vector $\boldsymbol{\lambda}$ and considering $\mathbf{f}^\mathrm{T}=\boldsymbol{\lambda}^\mathrm{T}\mathcal{M}$, we arrive at Bell inequalities,
        \begin{equation}
            \boldsymbol{\lambda }^{\mathrm{T}}\cdot\mathbf{P}\leq \sup\left(\boldsymbol{%
            \lambda }^{\mathrm{T}}\mathcal{M}\right),  \label{Eq:BellGeneral}
        \end{equation}
    which are in general not tight \cite{Brunner2014}.
    Violation of equation (\ref{Eq:BellGeneral}) implies the breaking local realism for a given set of the probability distributions $\mathbf{P}$.
    Search for the corresponding $\boldsymbol{\lambda }$ is a standard task of the aforementioned problem of linear programming \cite{Brunner2014}.


\section{Phase-space classical simulations}
\label{Sec:NonclProbl}

	In this section we establish a mathematical background for the problem of phase-space classical simulations.
	We demonstrate that the impossibility of such simulations is a sufficient condition for nonclassicality of quantum states.
	The main idea consists in simulating the outcomes of quantum measurements  with a classical device given the probability distributions $\mathcal{P}(A|a)$.
	Herewith, we will follow the general scheme explained in figure \ref{Fig:ClassState} assuming that the phase-space complex variables $\alpha$ play the role of the parameters $\omega$.
	For simplicity, we restrict ourselves by a single mode.
    
    According to Born's rule, the probability distributions $\mathcal{P}(A|a)$ are determined as 
    	\begin{equation}
    		\mathcal{P}(A|a)=\Tr\left[ \hat{\Pi}\left( A|a\right) \hat{\rho}\right],
    	\end{equation}
    where $\hat{\Pi}\left( A|a\right)$ are elements of the corresponding POVM. 
    In the phase-space representation, this rule for a single optical mode reads \cite{cahill69,cahill69a}
        \begin{equation}
            \mathcal{P}(A|a)=\int_{\mathbb{C}}\D^{2}\alpha \,\Pi \left( A|a;\alpha ;-s\right)
            P(\alpha ;s).  \label{Eq:EPRB_PS}
        \end{equation}%
    Here 
    	\begin{equation}\label{Eq:QuazPr}
    		P(\alpha ;s)=\Tr\left[ \hat{\rho}\hat{P}(\alpha ;s)\right]
    	\end{equation}
    and
    	\begin{equation}
    		\Pi\left( A|a;\alpha ;-s\right) =\frac{1}{\pi}\Tr\left[ \hat{\Pi}(A|a)\hat{P}(\alpha;-s)\right]
    	\end{equation}
    are the $s$-parameterized quasiprobability distribution and the $s$-parameterized POVM symbols, respectively.
    The operator $\hat{P}(\alpha;s)$ is given by 
    	\begin{equation}
    		\hat{P}(\alpha;s)=
    		\frac{2}{\pi(1-s)}:\exp\left[-\frac{2}{1-s}(\hat{a}^\dag-\alpha^\ast)(\hat{a}-\alpha)\right]:,
    	\end{equation}
    where $\hat{a}$ and $\hat{a}^{\dag }$ are boson field operators and $:\ldots:$ means the normal ordering.

    Observe that equation (\ref{Eq:EPRB_PS}) can be interpreted as a special form of equation (\ref{Eq:EPRB}), where the phase-space variables $\alpha$ play the role of the parameters $\omega$, i.e. $\Omega =\mathbb{C}$, and the POVM symbols are the response functions, $\Pi \left( A|a;\omega \right) =\Pi \left( A|a;\alpha ;-s\right)$.
    Indeed, for each measurement procedure, there always exists such $s$ that $\Pi \left( A|a;\alpha ;-s\right)\geq0$.
    For given $\mathcal{P}(A|a)$ and $\Pi \left( A|a;\alpha ;-s\right)$, equation (\ref{Eq:EPRB_PS}) can be considered as an integral equation with respect to unknown function $P(\alpha;s)$.
    Classical simulations are feasible if at least one solution to this equation is non-negative.
    We refer to such a procedure as phase-space classical simulations.
	We stress that the corresponding quasiprobability distribution is not necessarily a unique solution to equation (\ref{Eq:EPRB_PS}) if the measurement is not informationally \cite{busch91,DAriano2004,prugovecki77,Schroeck_book,renes2004} or tomographically complete.
	
	Let us now consider situations where the phase-space classical simulations are  not possible.
	This impossibility is a sufficient condition for appearance of negativities in phase-space quasiprobability distributions, i.e. nonclassicality of quantum states.
	However, this is not a necessary condition, since the solutions to the integral equation (\ref{Eq:EPRB_PS}) may not be unique for informationally and tomographically incomplete measurements.
	For instance, nonclassicality of quantum states does not necessarily means nonclassicality of photocounting statistics, since  photocounting experiments are not informationally complete.
	
	The impossibility of phase-space classical simulations for $s=1$ (the Glauber-Sudarshan representation) correspond to nonclassicality of quantum states motivated by equation (\ref{Eq:PFunc}).
	In this case, the corresponding POVM symbols,
	\begin{equation}
			\Pi\left( A|a;\alpha ;(-s)=-1\right)=\bra{\alpha}\hat{\Pi}\left( A|a\right)\ket{\alpha},
		\end{equation}
	are the probability distribution for the outcome $A$ given the setting $a$ and the coherent state $\ket{\alpha}$.
	This means that $\Pi\left( A|a;\alpha ;(-s)=-1\right)$ can be directly reconstructed in experiments by using a reference set of coherent states $\ket{\alpha}$.
	The feasibility of such reconstructions is a special feature of the Glauber-Sudarshan representation, which distinguishes it from other representations with $s<1$.

	If we focus on a given measurement scheme, the case of $s=1$ is not, in general, informative.
	This suggests to consider possibilities of phase-space classical simulations with other values of $s$.
	For example, P functions of Gaussian states may not be positive semidefinite.
	However, there exist other values of the parameter $s$ for which $P(\alpha;s)\geq0$; for example, the case of Wigner function $s=0$ or the case of Husimi-Kano Q function $s=-1$.
	If the POVM symbols $\Pi(A|a;\alpha;-s)$ of applied measurements are also non-negative, as e.g. for balanced homodyne detection and for eight-port homodyne detection (heterodyne detection) for $s=0$ and $s=-1$, respectively, then phase-space classical simulations are possible for the considered measurements\footnote{We stress that such simulations are possible with a general classical device. They are impossible with classical optical systems, where only coherent states and their statistical mixtures are employed.}; cf. references \cite{bartlett02,mari2012}.

	The POVM symbols $\Pi\left( A|a;\alpha ;-s\right)$ are always non-negative for $s=1$, and can also be non-negative for other values of $s<1$.
	For a given measurement, there exists a threshold value $s_\mathrm{th}$ such that $\Pi(A|a;\alpha;-s_\mathrm{th})\geq 0$ and $\Pi(A|a;\alpha;-s)\ngeq 0$ for $s<s_\mathrm{th}$.
	Let us assume that phase-space classical simulations are impossible for $s=s_\mathrm{th}$, i.e. non-negative solutions to equation (\ref{Eq:EPRB_PS}) for this value of $s$ do not exist.
	Since $\Pi(A|a;\alpha;-s)$ and $\Pi(A|a;\alpha;-s_\mathrm{th})$ are related through a convolution  \cite{cahill69,cahill69a}, we obtain that every solution to equation (\ref{Eq:EPRB_PS}) with $s>s_\mathrm{th}$ is uniquely related to a solution with $s=s_\mathrm{th}$ according to
		\begin{equation}
			P(\alpha;s_\mathrm{th})=\frac{2}{\pi(s-s_\mathrm{th})}\int_\mathbb{C}\D^2\gamma P(\gamma;s)\exp\left(-\frac{2|\alpha-\gamma|^2}{s-s_\mathrm{th}}\right).
			\label{Eq:Convolution}
		\end{equation}
	The above relation has exactly the same form as for the standard Cahill-Glauber quasiprobability distributions.
	Since $P(\alpha;s_\mathrm{th})\ngeq 0$ one gets $P(\alpha;s)\ngeq 0$.
	Thus, if phase-space classical simulations are impossible for $s=s_\mathrm{th}$, then they are also impossible for $s\geq s_\mathrm{th}$ and, as a consequence, for all values of $s\in[-1,1]$.

	The impossibility of phase-space classical simulations with all possible values of $s$ play a crucial role in quantum-simulation experiments, e.g. BosonSampling \cite{aaronson2013,scheel2004,scheel2008}.
	A related issue was discussed in reference \cite{rahimi-keshari16}. 
	Indeed, if equation (\ref{Eq:EPRB_PS}) applied to the corresponding multimode locally-operated scheme consists of non-positive constituents, it is not suitable for classical simulations.
	In principle, equation (\ref{Eq:EPRB_PS}) can be transformed into a relation with positive constituents only.
	However, computational hardness of such a procedure in general drastically increases with growing the number of modes.
	For example, finding such a transformation for standard BosonSampling schemes requires calculation of matrix permanents which is \#P-hard complexity class of calculations.


\section{Inequalities for phase-space classical simulations}
\label{Sec:DD-BI}
	
   Our treatment enables to formulate the dual form for the problem of phase-space classical simulations in the same way as it is performed for testing local realism.
   Specifically, we observe that equation~(\ref{Eq:EPRB_PS}) is similar to equation~(\ref{Eq:Marginal}), where $P(\alpha ;s)$ and $\Pi \left( A|a;\alpha ;-s\right)$ correspond to $\mathbf{W}$ and $\mathcal{M}$, respectively.
   Here summation is replaced by integration with respect to $\alpha$. 
   This yields inequalities for phase-space classical simulations,
        \begin{equation}
            \sum\limits_{a\in \mathcal{O}}E(\lambda |a){\leq }\sup_{\alpha \in
            \mathbb{C}}\sum\limits_{a\in \mathcal{O}}E(\lambda |a;\alpha ;s),
            \label{Eq:DDBell}
        \end{equation}
    where
        \begin{eqnarray}
	        &E(\lambda |a)=\sum_{A\in \mathcal{I}}\lambda(A,a)\mathcal{P}\left( A|a\right),\label{Eq:Expect1}\\
	        &E(\lambda |a;\alpha;s)=\sum_{A\in \mathcal{I}}\lambda(A,a)\Pi\left( A|a;\alpha;-s\right).\label{Eq:Expect2}
        \end{eqnarray}
    Here the sums may turn to integrals depending on the sets $\mathcal{O}$ and $ \mathcal{I}$. 
    The expectation values $E(\lambda |a)$ can be estimated experimentally from $M$ measured values, $A_i$, as 
        \begin{equation}
            E(\lambda |a)\approx\frac{1}{M}\sum_{i=1}^{M}\lambda (A_{i},a).
        \end{equation}
    If there exists $\lambda(A,a)$ that the inequality (\ref{Eq:DDBell}) is violated then phase-space classical simulations are impossible and, as a consequence, the state is nonclassical.
    In contrast to standard Bell inequalities (\ref{Eq:BellGeneral}), inequalities (\ref{Eq:DDBell}) carry information about the properties of the measurement device and the representation encoded in the POVM symbols $\Pi \left( A|a;\alpha;-s\right) $.
    These inequalities can be considered as a generalization of the result reported in reference \cite{rivas2009}.
	
	It is worth noting that the function $E(\lambda |a;\alpha ;s=1)$ appearing in the right-hand-side of equation (\ref{Eq:DDBell}) in the Glauber-Sudarshan representation can be directly reconstructed with a set of reference coherent states.
	Such a procedure is impossible for any other $s<1$ since the corresponding reference states would not be physical.
	This fact stresses a special role of the Glauber-Sudarshan representation in the general treatment of the notion of nonclassicality.
	
	Similar to standard Bell inequalities, there exists a set (in general, infinite) of test functions $\lambda$ corresponding to the tight inequalities for phase-space classical simulations.
	Such simulations of a given $\mathcal{P}(A|a)$ are impossible iff at least one inequality from this set is violated.
	Determining such an optimal set is in general a computationally hard problem.
	However, for testing such violations it is sufficient to find an example of $\lambda$ violating the corresponding inequality.


\section{Applications to typical measurement procedures}
\label{Sec:Appl}

	In this section we will apply inequalities (\ref{Eq:DDBell}) to some emblematic optical measurements.
	First, we will demonstrate that our approach enables to reveal nonclassicality of photocounting statistics even for super-Poissonian light.
	The device dependence enables us to apply this method inclusively to scenarios of realistic resolution between adjacent numbers of photons.
	It will be proven, however, that nonclassicality of photocounting statistics cannot be tested with a single on/off detector.
	Second, we employ our technique to unbalanced, balanced, and eight-port homodyne detections.
	These cases correspond to different threshold values $s_\mathrm{th}$ of the ordering parameter $s$.
	We will show that nonclassicality in these cases can be tested with tomographically incomplete sets of data.


\subsection{Nonclassical statistics of photocounts}
\label{Sec:Photcounting}

    Photoelectrical detection of light is a basic measurement in quantum optics.
    In this case, only the POVM symbols $\Pi _{Q}(n|\alpha )=\Pi(n|\alpha ;(-s){=-}1)$ are non-negative.
    Hence, only the case of $s=s_\mathrm{th}=1$ is relevant to this scenario.
    This corresponds to testing nonclassicality of quantum states associated with the P function or, to be more specific, to testing nonclassical statistics of photocounts.
    In addition, the sums in equation (\ref{Eq:DDBell}) contain only one term since the standard configuration of this experiment does not assume any setting.

    Here we consider ideal photon-number-resolving (PNR) detectors characterized by the POVM symbols \cite{kelley64,mandel_book},
        \begin{equation}
            \Pi _{Q}(n|\alpha )=\frac{|\alpha |^{2n}}{n!}\exp (-|\alpha |^{2}),\label{POVM:Mandel}
        \end{equation}
    where $A=n$ and $\mathcal{I}=\mathbb{N}$. 
    In addition, we analyse realistic photon-number resolution with click detectors, where the beam is split into $N$ spatial or temporal modes, each analysed by an on/off detector \cite{silberhorn2007,paul1996,castelletto2007,schettini2007,blanchet08,achilles03,fitch03,rehacek03}.
    The corresponding POVM symbols \cite{sperling12a} are given by
        \begin{equation}
            \Pi _{Q}(n|\alpha )={N\choose n}\left( 1-e^{-|\alpha |^{2}/N}\right)
            ^{n}e^{-|\alpha |^{2}(N-n)/N}. \label{Eq:POVM_Array}
        \end{equation}
    Here $\mathcal{I}=\{n\in \mathbb{N}|0\leq n\leq N\}$ and $A=n$ is the number of triggered detectors.
   	For these and all following examples, the detection efficiency $\eta\in[0,1]$ is attributed to quantum states and thus is not included in the POVMs.

    Nonclassical statistics of photocounts are commonly related to sub-Poissonian \cite{mandel79,mandel_book} or sub-binomial \cite{sperling12c,bartley13} ones for ideal and click detectors, respectively.
    However, such simple criteria do not always verify the presence of nonclassicality. 
    Typical examples are squeezed vacuum states 
    	\begin{equation}
    		\ket{r} =(\cosh r)^{-1/2}\sum_{n=0}^{+\infty }{2n\choose n}^{1/2}\left( \frac{\tanh r}{2}\right)^{n}\left\vert 2n\right\rangle,
    		\label{Eq:SVS}
    	\end{equation}
    where $r$ is the squeezing parameter, $\ket{n}$ are Fock number states, and even superpositions of coherent states, 
    	\begin{equation}
    		\ket{\psi} =\frac{1}{\sqrt{1+e^{-2|\alpha_{0}|^2}}}\Big(\ket{\alpha _{0}} +\ket{ -\alpha _{0}}\Big),
    		\label{Eq:SCS}
    	\end{equation}
    both attenuated with the efficiency $\eta$.
    For the latter case, nonclassical statistics can be tested with  ideal PNR detection using a particular form of inequalities (\ref{Eq:DDBell}), see reference \cite{rivas2009}. We show that an appropriate choice of $\lambda(n)$ enables to test nonclassicality in all considered situations including the case of imperfect photon-number resolution.

    Let us choose the test function $\lambda$ in equations (\ref{Eq:Expect1}) and (\ref{Eq:Expect2}) as  
    	\begin{equation}
    		\lambda(n)=(-t)^{n}e^{-gn^{2}},
    		\label{Eq:Lambda_PD}
    	\end{equation}
    where $t\geq-1$ and $g\geq0$.	
    For the ideal PNR detection we take $g=0$. 
    In this case, inequalities (\ref{Eq:DDBell}) are reduced to  
    	\begin{equation}
    		\frac{1}{\sqrt{1-[(1-\eta-t\eta)^2-1]\sinh^2r}}\leq1
    		\label{Eq:SVS_Ineq}
    	\end{equation}
    and 
    	\begin{equation}
    		\cosh [(1-\eta-t\eta)|\alpha _{0}|^{2}]\leq \cosh |\alpha _{0}|^{2}
    		\label{Eq:SCS_Ineq}
    	\end{equation}
    for the squeezed vacuum state and for the even superposition of coherent states, respectively. These inequalities are clearly violated for $t>(2-\eta)/\eta$. 
    We also note that the parameter $t$ in the case of squeezed vacuum states should be chosen such that $t<[1+\coth r-\eta]/\eta$; otherwise, the left-hand side of inequality (\ref{Eq:DDBell}) diverges.
    For click detectors with $N=10$, $\eta=0.6$ at $t=7$, $g=1/5$, the inequalities (\ref{Eq:DDBell}) are violated, giving $1.33\leq 1$ and $1.98\leq 1$ for the squeezed vacuum state, $r=0.7$, and for the even superposition of coherent states, $\alpha _{0}=1$, correspondingly. 
    This demonstrates that nonclassicality of photocounting statistics for these cases can be verified with click detectors.
    Details of the conducted algebra are given in \ref{App:PhotCount}.
	
    It directly follows from equation (\ref{Eq:DDBell}) that a single on/off detector, $N=1$, cannot be used for testing nonclassicality of photocounting statistics. 
    Indeed, the right-hand-side of inequality (\ref{Eq:DDBell}) is given in this case by
    	\begin{eqnarray}
			\sup_{\alpha\in\mathbb{C}}\Big[\lambda(0)\exp\left(-|\alpha|^2\right)+\lambda(1)&\left[1-\exp\left(-|\alpha|^2\right)\right]\Big]
			\nonumber\\
			&=\max\{\lambda(0),\lambda(1)\}.
		\end{eqnarray}
    At the same time, its left-hand side can be estimated as 
        \begin{equation}
          	\lambda(0)\mathcal{P}(0)+\lambda(1)\mathcal{P}(1)\leq\max\{\lambda(0),\lambda(1)\}.
		\end{equation}
    Hence, inequality (\ref{Eq:DDBell}) applied to on/off detectors is always fulfilled.
    Since direct and dual forms of the phase-space classical simulation problem are equivalent, we conclude that nonclassicality cannot be verified with a single on/off detector.
    A particular case of this property was discussed in reference \cite{sperling12c}.


\subsection{Unbalanced homodyne detection}
\label{Sec:UHD}

\begin{figure}[ht!!]
	    	\begin{center}
	    	\includegraphics[width=0.4\linewidth]{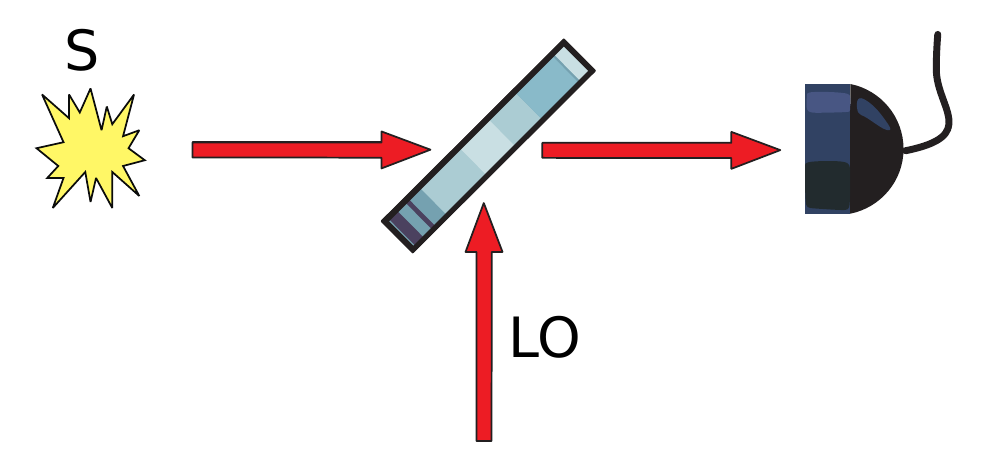}
	    	\end{center}
	    	\caption{\label{Fig:UHD} The scheme of UHD is depicted.
	    		The signal beam is superposed with the local oscillator on a beam splitter with the transmittance closed to 1 in order to perform a coherent displacement.
	    		The on/off detector analyzes presence or absence of photons in the field mode displaced in the phase space.}
    	\end{figure}

	Extension of the measurement scheme to unbalanced homodyne detection enables nonclassicality test with a single on/off detector.
	In this case, the signal mode interferes with a local oscillator in a beam splitter with a large transmittance, see figure \ref{Fig:UHD}.
	It results in a coherent displacement $\gamma\in\mathcal{O}$, see references \cite{wallentowitz96,mancini1997}, in the POVM symbols (\ref{Eq:POVM_Array}), $N=1$,
        \begin{equation}
            \Pi _{Q}(n|\gamma ;\alpha )=\left( 1-e^{-|\alpha -\gamma |^{2}}\right)
            ^{n}e^{-|\alpha -\gamma |^{2}(1-n)},  \label{Eq:POVM_UHD}
        \end{equation}
    where $A=n\in \mathcal{I}=\{0,1\}$. 
    In the most general case of $\mathcal{O}{=}\mathbb{C}$, this measurement is tomographically complete, i.e. the obtained information suffices for the reconstruction of the P function. 
    In practice, such a procedure is difficult to accomplish due to singularity of P functions. 
    
    Let us apply our method to the single-photon Fock state $\ket{1}$ attenuated with the efficiency $\eta=0.75$.
    Choosing the discrete settings $\mathcal{O}=\{\gamma _{1}=-0.1,\gamma_{2}=0,\gamma _{3}=0.1\}$ along with the test function $\lambda (0,\gamma_{1})=\lambda (0,\gamma _{3})=1$, $\lambda (0,\gamma _{2})=-2$, and $\lambda (1,\gamma_{k})=0$, we arrive at the violation of the inequality (\ref{Eq:DDBell}): $0.0099\leq 0.0089$. 
    In other words, nonclassicality of the single-photon state can be tested with three probabilities of no-click events.
    For details of calculations we reffer \ref{App:UHD}.


\subsection{Balanced homodyne detection}
\label{Sec:BHD}

    Impossibility of phase-space classical simulations for $s{\in}[0,1]$ and nonclassicality can also be analyzed with balanced homodyne detection (BHD). 
    In this case, the measurement of quadratures 
    	\begin{equation}
    		x(\varphi )=\frac{1}{\sqrt{2}}\left(\alpha e^{-i\varphi }+\alpha ^{\ast }e^{i\varphi }\right)\in \mathcal{I}=\mathbb{R},
    	\end{equation}
    cf. references  \cite{yuen1980,yuen1983,schumaker1984,yurke1987,schleich_book,VogelReview,vogel_book}, is described by the POVM symbols \cite{vogel_book, schleich_book},
 	    \begin{equation}
			\Pi\left(x|\varphi;\alpha;-s\right)=\frac{1}{\sqrt{\pi s}}\exp\left[-\frac{\left(x{-}\sqrt{2}\Re\alpha e^{-i\varphi}\right)^2}{s}\right].
			\label{Eq:POVM_BHD}
		\end{equation}
    For the continuous set of $\varphi \in \mathcal{O}=[0,\pi]$, this measurement is tomographically complete \cite{kvogel89, smithey93a}. 
    However, tests can also be conducted with discrete sets of quadratures, for instance choosing the phases as $\varphi_{k}=\pi k/K$ for $k=0\ldots K-1$, $K\in \mathbb{N}$.
    The impossibilities of phase-space classical simulations for BHD with $s=s_\mathrm{th}=0$ corresponding to negativities of the Wigner function \cite{Cartwright1976,Kenfack2004} leads to the impossibility for all $s\geq0$.

		\begin{figure}[th]
			\begin{center}
			\includegraphics[width=0.7\linewidth]{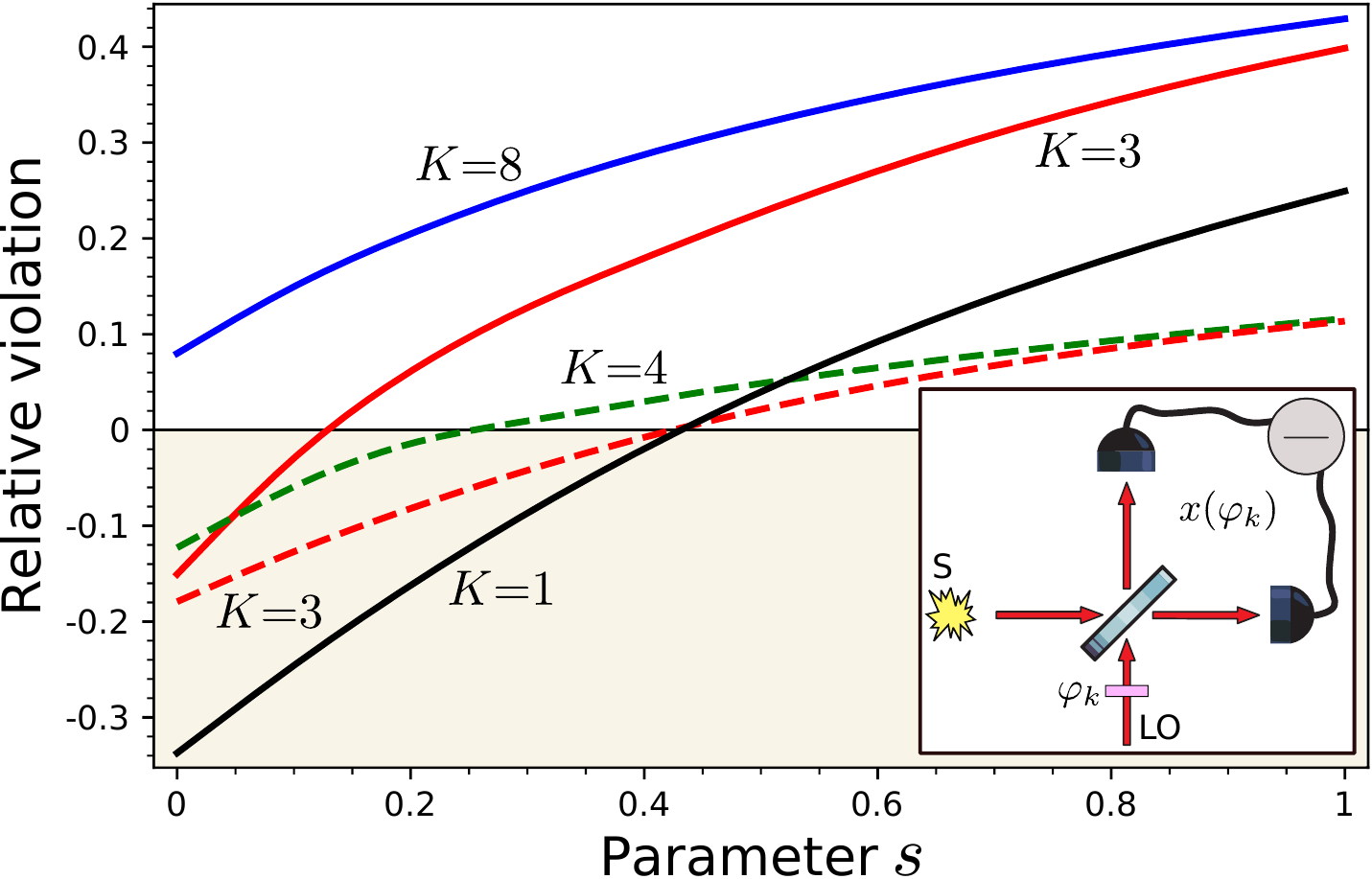}
			\end{center}
			\caption{Relative violation of inequalities (\ref{Eq:DDBell}) vs the parameter $s$ is shown for the Fock state $\left\vert 3\right\rangle $ in the scenario of BHD. 
				See the main text for details.
				The solid and dashed lines correspond to the efficiencies $\eta=1$ and $\eta=0.8$, respectively.
				The shaded area indicates the negative values, where the inequality is not violated. The inset sketches the scheme of BHD.}
			\label{Fig:BHD_violation}
		\end{figure}

    For example, for a three-photon state $\left\vert 3\right\rangle $, where the corresponding quandarature distribution is phase invariant,
    	\begin{equation}
    		\mathcal{P}(x)=(2^{3}3!\sqrt{\pi})^{-1}\Hermite_{3}^{2}(x)e^{-x^{2}},
    	\end{equation}
     we can pick the phase-independent test function 
     	\begin{equation}
     		\lambda (x,\varphi_{k})=\mathcal{P}(x).
     		\label{Eq:LambdaBHD}
     	\end{equation}
    Here $\Hermite_n(x)$ are Hermite polynomials.
    Similar strategy is applied to the case of the attenuated Fock state with the efficiency $\eta$. 
    In figure \ref{Fig:BHD_violation} we show the relative violation, i.e. the normalized difference of left- and right-hand sides of inequality (\ref{Eq:DDBell}), as a function of the parameter $s$. 
    One can observe that the presence of negativities can be tested with a finite number of settings $\varphi_k$ without reconstructing quasi-probability distributions.  
    Such a test can be performed for a wide range of $s$ with only one quadrature, in agreement with results of reference \cite{Shchukin_2004}. 
    For chosen $\lambda (x,\varphi _{k})$, negativity of the Wigner function, $s=0$, can be checked only for $K\geq 7$. 
    For $K=1,2$ negativity of the Wigner function is impossible to test in principle, since phase-space realistic models for these cases can be easily constructed.
    Details of the calculations are given in \ref{App:BHD}.


\subsection{Heterodyne and eight-port homodyne detection}
\label{Sec:8PHD}

    The measurement schemes of the complex field amplitudes $\alpha _{0}\in \mathcal{I}=\mathbb{C}$ with eight-port homodyne detection (EPHD) \cite{walker87} or heterodyne detection \cite{Helstrom_book} are described by the POVM symbols \cite{vogel_book, schleich_book},
        \begin{equation}
            \Pi (\alpha _{0}|\alpha ;-s)=\frac{2}{\pi (1+s)}\exp \left( -\frac{2}{1+s}%
            \left\vert \alpha _{0}-\alpha \right\vert ^{2}\right) .  \label{Eq:POVM_EPHD}
        \end{equation}
    These measurements have several noteworthy features:  
    (i) they have only one setting and, hence, the sums in equation (\ref{Eq:DDBell}) include only one term; (ii) they are informationally complete and $\mathcal{P}(\alpha_0)=Q(\alpha _{0})=P(\alpha_0;s=-1)\geq 0$ is the Husimi-Kano Q function \cite{husimi40,kano65}; 
    (iii) the POVM symbols (\ref{Eq:POVM_EPHD}) are non-negative for all $s{\in }[-1,1]$, i.e. $s_\mathrm{th}=-1$.
    Two later features mean that there always exist such a value of $s$ that the considered measurement can be classically simulated.
	In particular, such simulations are possible to perform by using the Husimi-Kano Q function.	
	However, a different choice of $s$ enables to test nonclassicality of quantum states with the eight-port homodyne detection (heterodyne detection).

       \begin{figure}[ht!!]
			\begin{center}
				\includegraphics[width=0.7\linewidth]{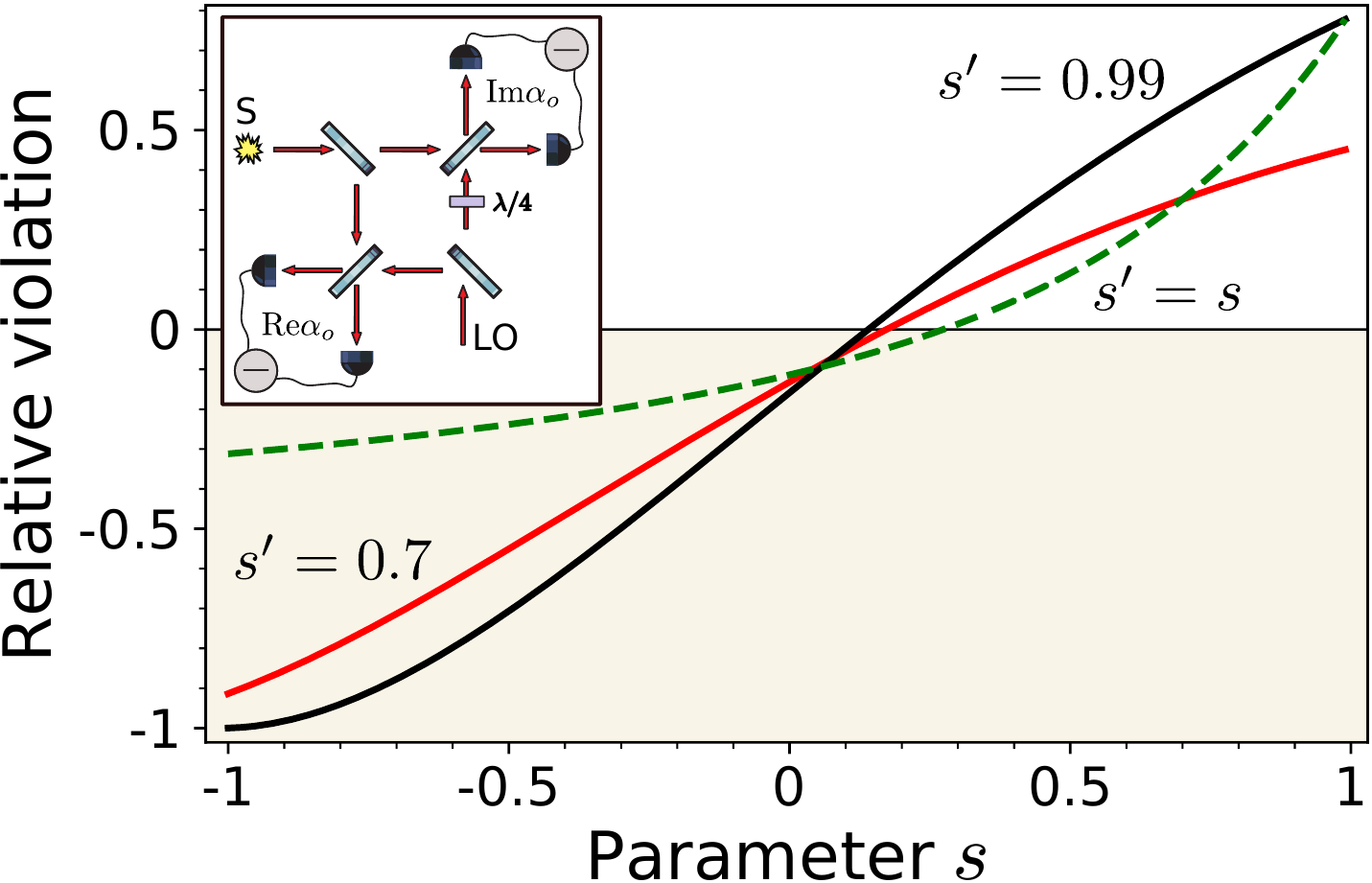}
			\end{center}
			\caption{Relative violation of inequality (\ref{Eq:DDBellEPHD2}) vs the parameter $s$ is shown for the Fock state $\left|1\right\rangle$ attenuated with the efficiency $\eta=0.8$ in the scenario of EPHD. 
			See the main text for details.
			The shaded area indicates negative values, where the inequality is not violated. The inset sketches the scheme of EPHD.}
			\label{Fig:EPHD_violation}
		\end{figure}
	
    Let us choose the parameter $\lambda (\alpha _{0})$ in the form of $s^{\prime }$-parameterized distribution, i.e. $\lambda (\alpha _{0})=P(\alpha_{0};s^{\prime })$. 
    Then, the inequality (\ref{Eq:DDBell}) reads as
        \begin{equation}
            \int_{\mathbb{C}}\D^{2}\alpha _{0}Q(\alpha _{0})P(\alpha _{0};s^{\prime
            })\leq \sup_{\alpha \in \mathbb{C}}P(\alpha ;s^{\prime }-s-1).
            \label{Eq:DDBellEPHD2}
        \end{equation}
    The above inequality can be considered as a necessary condition for positive semi-definiteness of $P(\alpha;s)$ expressed in terms of the Husimi-Kano Q function and other $s$-parameterized distributions. 
    The results of application of equation (\ref{Eq:DDBellEPHD2}) to the single-photon number state $\left\vert1\right\rangle$ attenuated with the efficiency $\eta=0.8$ are shown in figure~\ref{Fig:EPHD_violation}, where we plot the relative violation as a function of the parameter $s$. 
    One can observe that negativity of phase-space quasi-probability distributions for positive values of $s$, including the P function, can be tested in this scenario without their reconstruction.
    Details of calculations are presented in \ref{App:8PHD}.


\section{Conclusions}
\label{Sec:Conclusions}
	
	The classical simulations of quantum measurements are always possible if  no restrictions on the response functions are imposed and/or the computational hardness is not taken into account.
    A well-known example of restrictions on the response functions appear in the concept of local realism.
	The dual form of the corresponding problem leads to Bell inequalities, the violation of which implies the violation of local realism.
	
	We have considered another kind of restrictions in application to specific quantum optical systems.
	Constraining response functions to the form of phase-space POVM symbols enables to explore the idea of  phase-space classical simulations.
	A similar approach has been already considered in the particular context of BosonSampling schemes.
	Impossibility of phase-space classical simulation implies nonclassicality of quantum states.
	The dual form of this problem results in an analogue of Bell inequalities, which violation means nonclassicality of quantum states.
 
    Phase-space classical simulations is a device-dependent concept with a representation fixed by the ordering parameter $s$.
    For a given measurement procedure there always exists a threshold value $s_\mathrm{th}$ such that the impossibility of phase-space classical simulations for $s=s_\mathrm{th}$ leads to their impossibility for all other values of $s$.
    The threshold parameter $s_\mathrm{th}$ takes remarkable values -1, 0, and 1 for emblematic optical measurements: eight-port (heterodyne), balanced, and unbalanced homodyne detections, respectively.
    The latter case also includes the photocounting measurements.
    
    We have emphasized a special role of the Glauber-Sudarshan representation.
    In this representation, the POVM symbols and the right-hand side of inequalities for phase-space classical simulations can be directly reconstructed employing reference coherent states $\ket{\alpha}$.
	The reconstruction procedure is similar to sampling classical response functions with field modes of amplitudes $\alpha$.
	Such a direct sampling is impossible for other representations with $s<1$.
    
    We have demonstrated that our technique enables to test nonclassicality of quantum states in many practical situations.
    In particular, we have shown that nonclassicality of photocounting statistics is possible to test for quantum states and detection methods for which the standard approaches do not work.
	We have also demonstrated how to test nonclassicality of quantum states based on informationally and tomographically incomplete sets of measurement data.
	We believe that our results can be further applied in experimental research with quantum radiation fields.
    
    The authors thank S. Ryl, K. Pregracke, W. Vogel, V. Kovtoniuk, J. Sperling, and M. Bohmann for enlightening discussions. A.A.S. also acknowledges support from Department of Physics and Astronomy of the NAS of Ukraine through the Project 0120U101347. A.B.K acknowledges support from CONACyT (Mexico) Grant 254127.


\appendix


\section{Analysis of photocounting measurements}
\label{App:PhotCount}

    The calculations in the left-hand side of inequality (\ref{Eq:DDBell}) is convenient to conduct  in case of photocounting measurements, by including the efficiency $\eta$ in the POVM symbol, which is equivalent to its inclusion into quantum states.
    This implies the replacing $|\alpha|^2$ by $\eta|\alpha|^2$ in equations (\ref{POVM:Mandel}) and (\ref{Eq:POVM_Array}).
    The right-hand side of inequality (\ref{Eq:DDBell}) is still evaluated for the lossless POVM symbols since we consider losses to be included in quantum states.

\subsection{PNR detectors}

	In the case of ideal PNR detectors, the measurement outcomes are $A=n\in\mathcal{I}=\mathbb{N}$ and $\mathcal{O}$ consists of a single element.
	The POVM symbols, $\Pi_Q(n|\alpha)=\Pi(n|\alpha;s=-1)$, are given by equation~(\ref{POVM:Mandel}).
	Let us choose $\lambda(n)$ in the form of equation (\ref{Eq:Lambda_PD}) with $g=0$ and $t\geq-1$.
	The right-hand side of inequality~(\ref{Eq:DDBell}) is given by 
		\begin{eqnarray}
			\sup_{\alpha\in\mathbb{C}}\sum\limits_{n=0}^{+\infty}\lambda(n)\Pi_Q(n|\alpha)=&
			\sup_{\alpha\in\mathbb{C}}\sum\limits_{n=0}^{+\infty}(-t)^n\frac{|\alpha|^{2n}}{n!}\exp(-|\alpha|^2)
			\nonumber\\&=
			\sup_{\alpha\in\mathbb{C}}\exp\left[-|\alpha|^2(1+t)\right]=1.\label{Eq:PNR_RightSide}
		\end{eqnarray}
	The left-hand side of this inequality in this case can be expressed as a generating function of the photocounting distribution, 
		\begin{eqnarray}
			\sum\limits_{n=0}^{+\infty}\lambda(n)\mathcal{P}(n)=\sum\limits_{n=0}^{+\infty}(-t)^n\mathcal{P}(n)
			\nonumber\\
			=\sum\limits_{n=0}^{+\infty}(-t)^n\bra{\Phi}:\frac{(\eta\hat{n})^n}{n!}\exp[-\eta\hat{n}]:\ket{\Phi}
			\nonumber\\
			=\bra{\Phi}:\exp[-\eta(1+t)\hat{n}]:\ket{\Phi},
		\end{eqnarray}
	where $\eta$ is the efficiency and $\ket{\Phi}$ is the quantum state before suffering from losses.
	
	Consider the squeezed vacuum state, $\ket{\Phi}=\ket{r}$, given by equation (\ref{Eq:SVS}).
	In this case the left-hand side  of inequality~(\ref{Eq:DDBell}) is reduced to the form
		\begin{equation}
			\sum\limits_{n=0}^{+\infty}\lambda(n)\mathcal{P}(n)=
			\frac{1}{\sqrt{1-[(1-\eta-t\eta)^2-1]\sinh^2r}}.\label{Eq:SqVacLeftSide}
		\end{equation}
	Substituting equations~(\ref{Eq:PNR_RightSide}) and (\ref{Eq:SqVacLeftSide}) into inequality~(\ref{Eq:DDBell}) we get inequality (\ref{Eq:SVS_Ineq}).
	
	Consider the even superposition of coherent states, $\ket{\Phi}=\ket{\psi}$, given by equation (\ref{Eq:SCS}).
	The left-hand side of inequality~(\ref{Eq:DDBell}) in this case reads
		\begin{equation}
			\sum\limits_{n=0}^{+\infty}\lambda(n)\mathcal{P}(n)=
			\frac{\cosh[(1-t\eta-\eta)|\alpha _{0}|^{2}]}{\cosh|\alpha_0|^2}
			.\label{Eq:SchrCat_RightSide}
		\end{equation}	
	After substitution of equations (\ref{Eq:PNR_RightSide}) and (\ref{Eq:SchrCat_RightSide}) into inequality (\ref{Eq:DDBell}), the latter are reduced to inequality (\ref{Eq:SCS_Ineq}).
	
\subsection{Click detectors}

	For arrays of $N$ detectors the POVM symbols are given by equation~(\ref{Eq:POVM_Array}) where $|\alpha|^2$ is replaced with $\eta|\alpha|^2$.
	For practical purposes it is also useful to have an explicit form for its characteristic function,
		\begin{equation}
			\Pi_C(n|\beta;s)=
			\frac{1}{\pi^2}\int_\mathbb{C}\D^2\alpha \Pi(n|\alpha;s)\exp(\alpha^\ast\beta-\alpha\beta^\ast),\label{Eq:POVMChS}
		\end{equation}
	which reads
		\begin{eqnarray}
			\Pi_C(n|\beta;s)&=
	 		{N\choose n}
	 		\sum\limits_{k=0}^{n}
	 		{n\choose k}
	 		(-1)^{n-k}
			\frac{N}{\pi\eta(N-k)}
			\nonumber\\
			&\times\exp\left[-|\beta|^2\,\frac{(2-\eta)N+\eta k}{2\eta(N-k)}\right]
			\exp\left[s\frac{|\beta|^2}{2}\right].\label{Eq:ArrayDet_ChF}
		\end{eqnarray}	
	In this representation Born's rule for click statistics is given by
		\begin{equation}
			\mathcal{P}(n)=\int_\mathbb{C}\D^2\beta\, \Pi_C(n|\beta;-s)\, C(\beta;s),\label{Eq:BornRule_ChF}
		\end{equation}
	where
		\begin{equation}
			C(\beta;s)=\int_\mathbb{C}\D^2\alpha\, P(\alpha;s)\exp(\alpha^\ast\beta-\alpha\beta^\ast)
		\end{equation}	
	is the characteristic function of the Cahill-Glauber distribution.
	
	For the squeezed vacuum state (\ref{Eq:SVS}) the characteristic function for $s=0$ is given by
		\begin{eqnarray}
		 	C(\beta;s=0)
		 	\nonumber\\
		 	=\exp\left[-\frac{1}{4}
		 	\left(
		 	\begin{array}{cc}
		 		\beta^\ast & \beta
		 	\end{array}
		 	\right)
		 	\left(
		 	\begin{array}{cc}
			 	\cosh 2r & \sinh 2r\\
			 	\sinh 2r & \cosh 2r
		 	\end{array}
		 	\right)
		 	\left(
		 	\begin{array}{c}
			 	\beta\\
			 	\beta^\ast
		 	\end{array}
		 	\right)
		 	\right].\label{Eq:SqVac_ChF}
		\end{eqnarray}
	 Substituting equations (\ref{Eq:ArrayDet_ChF}) and (\ref{Eq:SqVac_ChF}) into equation (\ref{Eq:BornRule_ChF}), we get the probability distributions of clicks for the squeezed vacuum state,
	 	\begin{eqnarray}
	 		\mathcal{P}(n)=
	 		{N\choose n}
	 		&\sum\limits_{k=0}^{n}
	 		{n\choose k}
	 		(-1)^{n-k}
	 		\nonumber\\
	 		&\times\frac{N}{\sqrt{N^2+\eta[N-k][(2-\eta)N+\eta k]\sinh^2r}}.
	 	\end{eqnarray}
	For the even superposition of coherent states given by equation~(\ref{Eq:SCS}) this distribution reads
	 	\begin{eqnarray}
			\mathcal{P}(n)={N\choose n}
	 		&\sum\limits_{k=0}^{n}
	 		{n\choose k}
	 		(-1)^{n-k}\nonumber\\
			&\times\frac{\cosh\left[\left(\frac{\eta k}{N}+1-\eta\right)|\alpha_0|^2\right]}{\cosh|\alpha_0|^2}.
		\end{eqnarray}	
	
\begin{figure}[ht!!]
			\begin{center}
			\includegraphics[width=0.8\linewidth]{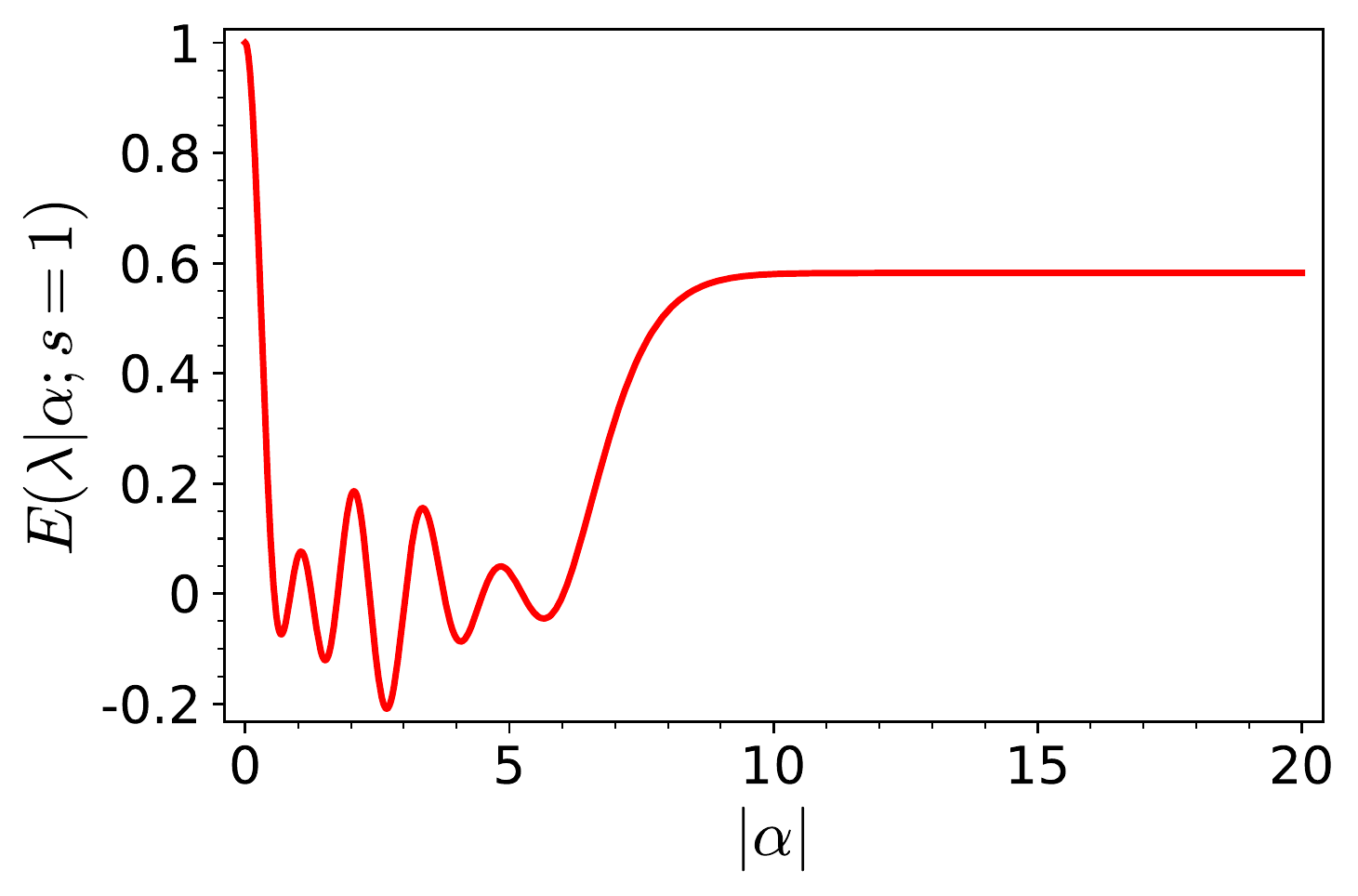}
			\end{center}
			\caption{\label{Fig:Array-Right} Dependence of $E(\lambda|\alpha;s=1)$ on $|\alpha|$ for click detectors with $N=10$ and for $\lambda(n)=(-t)^ne^{-g n^2}$, $t=7$, and $g=1/5$.
			The maximal value of $E(\lambda|\alpha;s=1)$ is 1 for $|\alpha|=0$.}
		\end{figure}

	We consider the case of $N=10$ and choose $\lambda(n)$ in the form of equation (\ref{Eq:Lambda_PD}) with $t=7$, $g=1/5$, and $\eta=0.6$.
	Evaluation of right- and left-hand sides of inequality~(\ref{Eq:DDBell}) was performed numerically.
	For this purpose we consider $r=0.7$ and $\alpha_0=1$ for squeezed vacuum state (\ref{Eq:SVS}) and even superposition of coherent states (\ref{Eq:SCS}), respectively.
	The right-hand side of inequality~(\ref{Eq:DDBell}) in this case can be evaluated from the plot presented in figure~\ref{Fig:Array-Right},
		\begin{equation}
			\sup_{\alpha\in\mathbb{C}}E(\lambda|\alpha;s=1)=
			\sup_{\alpha\in\mathbb{C}}\sum\limits_{n=0}^{N}\lambda(n)\Pi_Q(n|\alpha)=1,
		\end{equation}
	where we set $\eta=1$ in the POVM symbol since losses are considered to be attributed solely to the quantum states.
	The numerically-evaluated left-hand side is 1.33 and 1.98 for squeezed vacuum state (\ref{Eq:SVS}) and even superposition of coherent states (\ref{Eq:SCS}), correspondingly.
	Therefore, the inequalities are clearly violated.
		
\section{Analysis of unbalanced homodyne detection}
\label{App:UHD}

	The POVM symbols for UHD with a single on/off detector are given by equation (\ref{Eq:POVM_UHD}).
	We choose the discrete set of settings $\mathcal{O}=\{\gamma_1=-0.5,\gamma_2=0,\gamma_3=0.5\}$ and the test function as $\lambda(0,\gamma_1)=\lambda(0,\gamma_3)=1$, $\lambda(0,\gamma_2)=-2$, and $\lambda(1,\gamma_k)=0$.
	In this case the right-hand side of inequality~(\ref{Eq:DDBell}) is given by
		\begin{equation}
			\sup_{\alpha\in\mathbb{C}}\sum_{\gamma_k\in\mathcal{O}}E(\lambda|\gamma_k;\alpha;s=1)
			\approx 0.0089.\label{Eq:UHD_RightSide}
		\end{equation}

    We apply our method to the single-photon Fock state $\ket{1}$, which after attenuation with an efficiency $\eta$ takes the form
		\begin{equation}
			\hat{\rho}=\eta\ket{1}\bra{1}+(1-\eta)\ket{0}\bra{0}.
		\end{equation}
	The no-click POVM element for UHD can be rewritten in the form
		\begin{equation}
			\hat{\Pi}(0|\gamma_k)=\ket{\gamma_k}\bra{\gamma_k}.
		\end{equation}
	Then the probability of no click event reads
		\begin{eqnarray}
			\mathcal{P}(0|\gamma_k)&=\eta\big|\left\langle\gamma_k|1 \right\rangle\big|^2+(1-\eta)\big|\left\langle\gamma_k|0 \right\rangle\big|^2
			\nonumber\\
			&=\left(\eta|\gamma_k|^2+1-\eta\right)\exp\left(-|\gamma_k|^2\right).
		\end{eqnarray}
	This expression is used for evaluation of the left-hand side of inequality~(\ref{Eq:DDBell}),
		\begin{equation}
			\sum_{\gamma_k\in\mathcal{O}}E(\lambda|\gamma_k)=
			\sum_{\gamma_k\in\mathcal{O}}\sum_{n=0}^{1}\lambda(n,\gamma_k)\mathcal{P}(n|\gamma_k)
			\approx 0.0099.
		\end{equation}
	Comparing it with the right-hand side given by equation (\ref{Eq:UHD_RightSide}), we conclude that the inequality is violated.


\section{Analysis of balanced homodyne detection}
\label{App:BHD}

	The POVM for BHD is represented by the projectors on eigenstates of the quadrature operator, $\hat{\Pi}(x|\varphi)=\ket{x(\varphi)}\bra{x(\varphi)}$.
	The POVM symbols are given by equation (\ref{Eq:POVM_BHD}).
	The corresponding measurement outcomes $A=x\in\mathcal{I}=\mathbb{R}$ take continuous values.
	We consider a discrete set of settings, $\varphi_k=\pi k/K$ for $k=0\ldots K-1$, where $K\in\mathbb{N}$.

	We choose $\lambda(x|\varphi_k)$ in the form of the quadrature distribution of the Fock state $\ket{n}$, cf. equation (\ref{Eq:LambdaBHD}), considering also the case of an arbitrary $n$ for which
		\begin{equation}
			\lambda(x|\varphi_k)=\mathcal{P}_n(x)=\frac{1}{2^nn!\sqrt{\pi}}
			\Hermite_n^2\left(x\right)\exp\left(-x^2\right),
		\end{equation}
	where $\Hermite_n(x)$ is the Hermite polynomial.
	This function does not depend on $\varphi_k$, which is motivated by its application to phase-independent states.
	In this case we have 
		\begin{eqnarray}
			&E_n(\lambda|\varphi_k;\alpha;s)=\int_\mathbb{R}\D x \lambda(x|\varphi_k)\Pi\left(x|\varphi_k;\alpha;-s\right)\frac{1}{2^nn!\sqrt{\pi(s+1)}}\label{Eq:BHD_RightSideEta1}\\
			&\times\sum\limits_{l=0}^{n}2^ll!{n\choose l}^2\left(\frac{1}{s+1}\right)^{n-l}\Hermite_{2(n-l)}\left(\frac{x_0(\varphi_k)}{\sqrt{s+1}}\right)\exp\left[-\frac{x_0^2(\varphi_k)}{s+1}\right],
			\nonumber
		\end{eqnarray}
	where $x_0(\varphi_k)=2^{-1/2}(\alpha e^{-i\varphi_k}+\alpha^\ast e^{i\varphi_k})$.	
	The right-hand side of inequality~(\ref{Eq:DDBell}),
		\begin{equation}
			\sup_{\alpha\in\mathbb{C}}\sum\limits_{k=0}^{K-1}E_n(\lambda|\varphi_k;\alpha;s),
		\end{equation}
	can be numerically estimated.
	Here the presence of multiple local maxima should be taken into account. 
	
	Let us consider the Fock states, $\ket{n}$.
	In this case, the left-hand side of inequality~(\ref{Eq:DDBell}) reads
		\begin{eqnarray}
			&\sum\limits_{k=0}^{K-1}E(\lambda|\varphi_k)
			=KE(\lambda|\varphi_k)=
			K\int_\mathbb{R}\D x\mathcal{P}^2(x)\\
			&=\frac{K(n!)^2}{2^{2n}\sqrt{\pi}}
			\sum\limits_{l=1-(-1)^n}^{2n}2^{l+1/2}l!
			{{-1/2}\choose l}
			\nonumber\\
			&\times\sum\limits_{m=\max\left[\frac{1-(-1)^n}{2},l-n\right]}^{\min\left[n,l-\frac{1-(-1)^n}{2}\right]} A(n,m)A(n,l-m),
			\nonumber
		\end{eqnarray}
	where
		\begin{eqnarray}
			&A(n,m)\label{Eq:CoeffA}
			\\
			&=\sum\limits_{i=\max\left[0,\lceil \frac{n}{2}\rceil-m\right]}^{\min\left[\lfloor\frac{n}{2}\rfloor,n-m\right]}
			\frac{1}{i!(n-i-m)!(n-2i)!(2i+2m-n)!}.\nonumber
		\end{eqnarray}
	In subsection \ref{Sec:BHD} we have analyzed the case of $n=3$.
	The other Fock states can also be tested with this technique.
	
	In the case of attenuated states, the probability distribution $\mathcal{P}_n(x)$ is modified according to
		\begin{equation}
			\mathcal{P}_n(x;\eta)=\sum\limits_{m=0}^{n}{n\choose m}\eta^m(1-\eta)^{n-m}\mathcal{P}_m(x).
		\end{equation}	
	It can also be used as the test function for the attenuated state.
	Therefore, the expression under supremum at the right-hand side of inequality (\ref{Eq:DDBell}) is given by
		\begin{equation}
			E(\lambda|\varphi_k;\alpha;s)=\sum\limits_{m=0}^{n}{n\choose m}\eta^m(1-\eta)^{n-m}E_m(\lambda|\varphi_k;\alpha;s),
		\end{equation}
	where $E_m(\lambda|\varphi_k;\alpha;s)$	is given by Eq. (\ref{Eq:BHD_RightSideEta1}).
	The left-hand side of inequality (\ref{Eq:DDBell}) in this case reads
		\begin{eqnarray}
			\sum\limits_{k=0}^{K-1}E(\lambda|\varphi_k)=KE(\lambda|\varphi_k)\\
			=\sum\limits_{m_1=0}^{n}\sum\limits_{m_2=0}^{n}{n\choose {m_1}}{n\choose{m_2}}\eta^{m_1+m_2}
			(1-\eta)^{2n-m_1-m_2}I_{m_1m_2},\nonumber
		\end{eqnarray}	
	where
		\begin{eqnarray}
			I_{m_1m_2}=\frac{m_1!m_2!(-1)^{m_1+m_2}}{2^{m_1+m_2}\sqrt{2\pi}}
			\\
			\times\sum_{k=\frac{1-(-1)^{m_1}}{2}}^{m_1}\sum_{l=\frac{1-(-1)^{m_2}}{2}}^{m_2}
			2^{k+l}{{-1/2}\choose{k+l}}(k+l)!A(m_1,k)A(m_2,l),\nonumber
		\end{eqnarray} 
	and $A(m,k)$ is given by equation (\ref{Eq:CoeffA}).


\section{Analysis of eight-port homodyne detection and heterodyne detection}
\label{App:8PHD}

	Eight-port homodyne detection and heterodyne detection are described with the POVM in form of the projectors on coherent states, $\hat{\Pi}(\alpha_0)=\pi^{-1}\ket{\alpha_0}\bra{\alpha_0}$.
	The POVM symbols of these measurement procedures are given by equation (\ref{Eq:POVM_EPHD}). 
	In this case, right- and left-hand side of inequality (\ref{Eq:DDBell}) are expressed in terms of
	 	\begin{equation}
	 		E(\lambda|\alpha;s)=\int_\mathbb{C}\D^2\alpha_0 \lambda(\alpha_0)\Pi(\alpha_0|\alpha;-s)
	 		\label{Eq:DDBell_8PHD_RS}
	 	\end{equation}
	and
		\begin{equation}
			E(\lambda)=\int_\mathbb{C}\D^2\alpha_0 \lambda(\alpha_0)Q(\alpha_0),
		\end{equation}
	respectively.
	Here $Q(\alpha_0)=\mathcal{P}(\alpha_0)=P(\alpha_0;s=-1)$ is the Husimi-Kano Q function.
	Let us choose $\lambda(\alpha_0)=P(\alpha_0;s^\prime)$.
	Using the convolution formula similar to equation (\ref{Eq:Convolution})one gets for equation (\ref{Eq:DDBell_8PHD_RS})
		\begin{equation}
			E(\lambda|\alpha;s)=P(\alpha;s^\prime-s-1).	
		\end{equation}

	As an example, we consider the Fock state $\ket{1}$ attenuated with the efficiency $\eta$.
	The corresponding $s$-parameterized quasi-probability distribution reads
		\begin{equation}
			P(\alpha;s)=\frac{2}{\pi(1-s)^2}\left(\frac{4\eta|\alpha|^2}{1-s}-2\eta+1-s\right)
			\exp\left(-\frac{2|\alpha|^2}{1-s}\right).
		\end{equation}
	This yields right- and left-hand sides of DD Bell inequalities, 
		\begin{eqnarray}
			\sup_{\alpha\in\mathbb{C}}E(\lambda|\alpha;s)
			\\
			=\frac{2}{\pi(2+s-s^\prime)}
			\left\{
			\begin{array}{lcr}
				2\eta\exp\left(-\frac{4\eta-2+s^\prime-s}{2\eta}\right)&\textrm{for}&s\leq4\eta-2+s^\prime\\[3ex]
				2-2\eta+s-s^\prime&\textrm{for}&s\geq 4\eta-2+s^\prime
			\end{array}
			\right.\nonumber
		\end{eqnarray}
	and
	 	\begin{eqnarray}
			E(\lambda|\alpha;s)&=\frac{16\eta^2}{\pi(3-s^\prime)^3}
			\\
			&+\frac{2}{\pi(3-s^\prime)^2}
			\left[\frac{4\eta(1-\eta)}{1-s^\prime}+\eta(1-2\eta-s^\prime)\right]
			\nonumber
			\\
			&+\frac{2(1-\eta)(1-2\eta-s^\prime)}{\pi(1-s^\prime)(3-s^\prime)},
			\nonumber
		\end{eqnarray}	
	respectively.

\section*{References}
\bibliography{biblio}

\end{document}